# $H_\infty$ Controller Design based on the T-S Triangular Cloud Model

UnSun Pak, YongNam Kim, GyongIl Ryang

Faculty of Electronics & Automation, **Kim Il Sung** University,

Pyongyang, Democratic People's Republic of Korea

**Abstract:** In this paper, we propose a design method for controller based it on that describe plants as T-S triangular cloud models in case of uncertainty in them.

**Keywords**: $H_\infty$ Controller, T-S Triangular Cloud Model, triangle cloud control

## 1. T-S Triangular Cloud Model

In previous work [1], an approach for $H_\infty$ controller design based on the normal model was proposed, but the drawback is it is necessary to calculate a lot and is difficult to analyze.
In previous work [2], a method for $H_\infty$ controller design in case the plant is described as an indefinite model was also proposed and the drawback of this method is that identify uncertainty of plant analytical.

**Definition 1.** Let T-S triangular cloud model is defined by the rule represented as the following form.

If $x_1$ is $A_1$ and $x_2$ is $A_2$, …, $x_n$ is $A_n$ then $y = a_0 + a_1 x_1 + \cdots + a_n x_n$, where $A_1, \ldots, A_n$ are triangular cloud models of language values and $a_0, \ldots, a_n$ are random values varied by degree of uncertainty of condition.

The following form shows how to identify T-S triangular cloud model with plant by the rule of if-then.

$L^l$ : if $x_1$ is $A_1^l$, …, $x_n$ is $A_n^l$ then $y^l = a_0^l + a_1^l x_1 + \cdots + a_n^l x_n$ (1), where $L^l$ $(l = 1, \cdots, r)$ is $l^{th}$ rule for plant, $x_i$ $(i = 1, \cdots, n)$ is input of it and $y^l$ is output of it.

The condition part of it has the same structure with degree $n$ of X condition cloud model and then, the output, which is obtained several values, not a single value, therefore reflects how uncertain the plant is. (Picture 1)

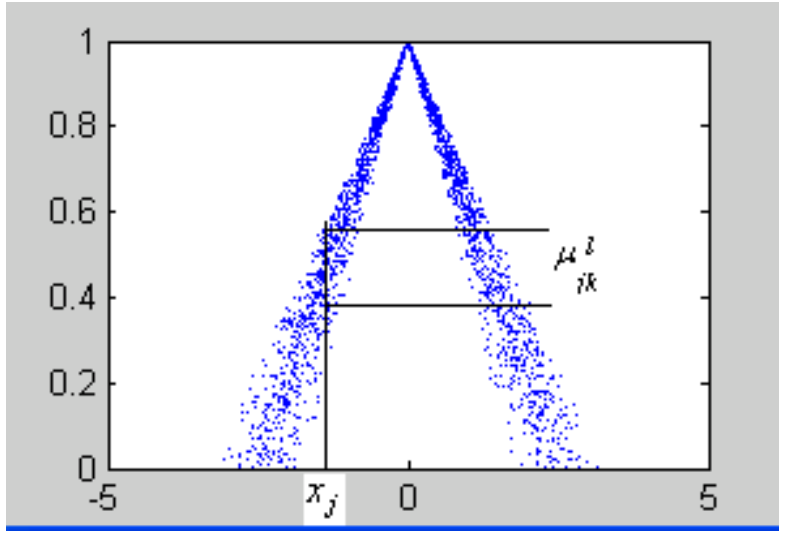

Picture 1. Precision according to the input of cloud model

To reflect its uncertainty on the conclusion, $a^l_j$ is defined by following.

$a^l_0 = \bar{a}^l_0$, $a^l_j = N(\bar{a}^l_j, \sigma^l_j)$, $(j = 1, \cdots, n)$, where

$\sigma^l_j = \max\{\mu^l_{jk}\} - \min\{\mu^l_{jk}\}$, $(k = 1,\ 2, \cdots)$.

And $\bar{a}^l_j$, where is parameter obtained by identification of conclusion, is the expected value of membership cloud. If all inputs are equal to the expected values of triangular cloud model, then $\sigma^l_j = 0$ and $a^l_j = \bar{a}^l_j$.

On the other hand, the output $y^0$ by *r* rules for input $x^0_1, \cdots, x^0_n$ is represented by the following.

$y^0 = \sum_{l=1}^{r} w^l \cdot y^l / \sum_{l=1}^{r} w^l$ _(2), where $w^l = \prod_{i=1}^{n} A^i_j(x^0_j)$, $y^l = a^l_0 + a^l_1 x_1 + \cdots + a^l_n x_n$ and $A_j(x^0_j)$ is first order *X* condition cloud model of $X^0_j$ for the expected curve of triangular cloud model $A_j$.

**Definition 2.** T-S triangular cloud model of dynamic system is represented by the following rule.

$$\dot{x}(t) = \frac{\sum_{l=1}^{r} w_l(t)[(A_l + \Delta A_l)x(t) + (B_l + \Delta B_l)u(t)]}{\sum_{i=1}^{r} w_i(t)} = \sum_{i=1}^{r} h_l(t)[(A_l + \Delta A_l)x(t) + (B_l + \Delta B_l)u(t)]$$

$$y(t) = \frac{\sum_{l=1}^{r} w_i(t)(C_l + \Delta C_l)x(t)}{\sum_{l=1}^{r} w_l(t)} = \sum_{l=1}^{r} h_l(t)(C_l + \Delta C_l)x(t) \qquad (3)$$

In the above equation, $A_l \in R^{n\times n}, B_l \in R^{n\times m}, C_l \in R^{s\times n}, x \in R^n, u \in R^m, and\ y \in R^p$ and $w_l(t)$ is drawn from the conditions of rule. And it is necessary that the following condition is satisfied at each instant *t*.

$$w_l(t) \ge 0, \quad \sum_{i=1}^{r} w_l(t) > 0, h_l(t) = \frac{w_l(t)}{\sum_{l=1}^{r} w_l(t)}, \quad \sum_{l=1}^{r} h_l(t) = 1$$

It follows from the definition of T-S triangular cloud model of the above dynamic system that the plant is identified by the below rules.

**Plant rule:**

If $Z_1(t)$ is $M_{l1}$ and … and $Z_p(t)$ is $M_{lp}$ , then

$$\begin{cases} \dot{x}(t) = (A_l + \Delta A_l)x(t) + (B_l + \Delta B_l)u(t) \\ y(t) = (C_l + \Delta C_l)x(t) \end{cases}$$

, where $l = \overline{1, r}$ is the number of rules, $Z_j(t)$ is the input of rule, $M_{ij}$ is the language value of triangular cloud model and $\Delta A_l$, $\Delta B_l$, $\Delta C_l$ is relative to the width of plant cloud model as the uncertain part.

The plant cloud model for the rule $l$ is represented by the following.

$$\begin{cases} \dot{x}(t) = \sum_{l=1}^{r} h_l(Z(t)) \cdot \{(A_l + \Delta A_l)x(t) + (B_l + \Delta B_l)u(t)\} \\ y = \sum_{l=1}^{r} h_l(Z(t)) \cdot (C_l + \Delta C_l)x(t) \end{cases} \qquad (4)$$

## 2. An approach to design $H_\infty$ controller based on the T-S Triangular Cloud Model

An uncertain system discussed is represented by the following.

$$\begin{cases} \dot{x} = [A + \Delta A(t)]x + [B + \Delta B(t)]u \\ y = [C + \Delta C(t)]x \end{cases} \qquad (5),$$

$$x \in R^n, u \in R^m, y \in R^p, \Delta A(t) \in R^{n\times n}, \Delta B(t) \in R^{n\times m}, \Delta C(t) \in R^{p\times n}$$

For the plant described by the equation (5), the following supposition is introduced.

**[Supposition]**

① With constant matrices $D_{jl}, E_{jl}$ and time variant matrices $\Delta a_l(t), \Delta b_l(t), \Delta c_l(t)$,

$\Delta A_{l;}(t), \Delta B_l(t), \Delta C_l(t)$ are described by the following.

$$\Delta A_l = D_{1l}\Delta a_l(t)E_{1l} \qquad \Delta a_l(t) = \Delta a_l^T(t)$$
$$\Delta B_l = D_{2l}\Delta b_l(t)E_{2l} \qquad \Delta b_l(t) = \Delta b_l^T(t)$$
$$\Delta C_l = D_{3l}\Delta c_l(t)E_{3l} \qquad \Delta c_l(t) = \Delta c_l^T(t)$$

② Matrices $A_l, B_l, C_l, D_{1l}, D_{2l}, D_{3l}, E_{1l}, E_{2l}, E_{3l}$ are already known.

③ Matrices $\Delta a_l(t), \Delta b_l(t)$, $\Delta c_l(t)$ are not known but they're satisfied the equation

$\Delta a_l^T(t)\Delta a_l(t) < \mathrm{I}$ , $\Delta b_l^T(t)\Delta b_l(t) < \mathrm{I}$ , $\Delta c_l^T(t)\Delta c_l(t) < \mathrm{I}$ $(j = 1, 2, 3)$ .

④ The signal $y$ is possible to measure.

For the uncertain system (5) satisfied supposition ①~④, cloud model dynamic compensator is defined by the following. [2]

**Control rule**

$$\text{If } Z_1(t) \text{ is } M_{l1} \text{ and} \ldots \text{and } Z_p(t) \text{ is } M_{lp} \text{ then } \begin{cases} \dot{x}_c = \hat{A}_l x_c + \hat{B}_l y \\ u = \hat{C}_l x_c \end{cases}$$

Also, control rule is described by the following to represent all rules simple.

$$\begin{cases} \dot{x}_c = \sum_{l=1}^{r} h_l(Z(t)) \cdot (\hat{A}_l x_c + \hat{B}_l y) \\ u = \sum_{l=1}^{r} h_l(Z(t)) \cdot \hat{C}_l x_c \end{cases} \qquad (6)$$

**i. Margin of uncertainty (maximum width of cloud)**

In general, when cloud model $A(Ex, En, He)$ is known, most cloud drops are dropped down between the curves $y_1 = \begin{cases} 1 - \left|\dfrac{x - Ex}{En - 3He}\right|, & |x - Ex| \le |En - 3He| \\ 0, & \text{the other case} \end{cases}$ and

$y_2 = \begin{cases} 1 - \left|\dfrac{x - Ex}{En + 3He}\right|, & |x - Ex| \le |En + 3He| \\ 0, & \text{the other case} \end{cases}$. (Picture 2)

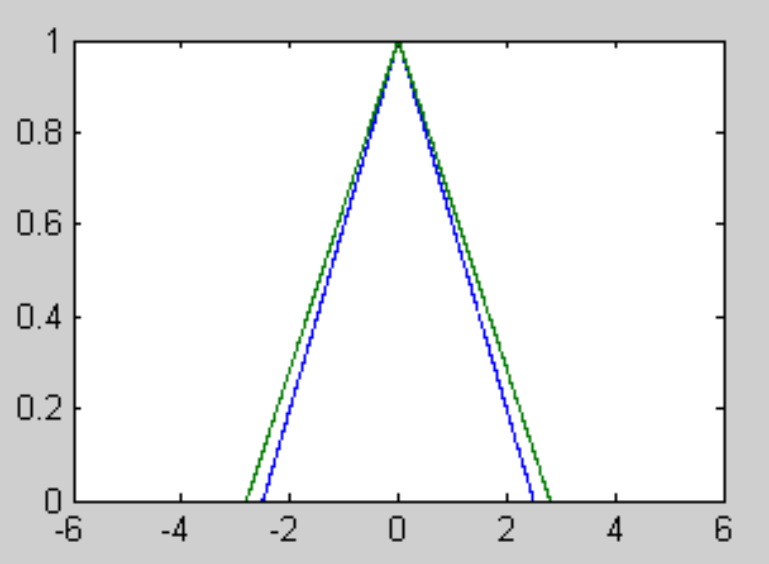

Therefore, width for one certain point $x$ is the difference between two curves.

$$d = y_1 - y_2 = 1 - \left|\frac{x - Ex}{En + 3He}\right| - 1 + \left|\frac{x - Ex}{En - 3He}\right| = \left|\frac{x - Ex}{En - 3He}\right| - \left|\frac{x - Ex}{En + 3He}\right|$$

At the point for maximum width,

$$d' = (y_1 - y_2)' = \left( \left| \frac{x - Ex}{En - 3He} \right| - \left| \frac{x - Ex}{En + 3He} \right| \right)' = \left| \frac{x - Ex}{En - 3He} \right|' - \left| \frac{x - Ex}{En + 3He} \right|' =$$

$$= \left( \frac{1}{|En - 3He|} - \frac{1}{|En + 3He|} \right) |x - Ex|'$$

$$= \begin{cases} \left( \frac{1}{|En - 3He|} - \frac{1}{|En + 3He|} \right)(x - Ex), & (x - Ex) > 0 \\ -\left( \frac{1}{|En - 3He|} - \frac{1}{|En + 3He|} \right)(Ex - x), & (x - Ex) < 0 \end{cases}$$

Because $x$ is null when $d' = 0$ according to the above equation, maximum width of cloud at the point $y_1 = 1/3$ is calculated using the characteristic of triangular membership cloud function by the following.

$$y_1 = \frac{1}{3} = 1 - \left| \frac{x - Ex}{En - 3He} \right|$$

$$|x - Ex| = \frac{2}{3} |En - 3He|$$

$$y_2 = 1 - \left| \frac{x - Ex}{En + 3He} \right| = 1 - \left| \frac{x - Ex}{En + 3He} \right| = 1 - \frac{|x - Ex|}{|En + 3He|} = 1 - \frac{\frac{2}{3}|En - 3He|}{|En + 3He|} = 1 - \frac{2}{3} \left| \frac{En - 3He}{En + 3He} \right|$$

$$d_{\max} = y_2 - y_1 = 1 - \frac{2}{3} \left| \frac{En - 3He}{En + 3He} \right| - \frac{1}{3}$$

After the triangular cloud model is known, its width is fixed as $d_{jl} < d_{\max} = d_l^0$.

So let maximum width $d_l^0$ is decided by the following. $d_l^0 = 1 - \frac{2}{3} \left| \frac{En - 3He}{En + 3He} \right| - \frac{1}{3}$ (7).

**ii. Design of $H_\infty$ controller based on T-S triangular cloud model**

Let the width adjusted to triangular cloud model $M_{l1}, \cdots, M_{lp}$ in the condition of the plant rule is called $d_{l1}, \cdots, d_{lp}$. Then, it follows that $d_{lj} \le d_{lj}^{\max}$, where $d_{lj}^{\max}$ is the maximum width of $M_{lj}$.

It follows from the equation $d_l^{\max} = \max_j d_{jl}^{\max}$ that obtains the equation

$$\|\Delta a_l\| < d_l^{\max}, \ \|\Delta b_l\| < d_l^{\max}, \ \|\Delta c_l\| < d_l^{\max} \quad (8).$$

Introduce the following symbols.

$$C_a \stackrel{\Delta}{=} [E_{11}^T \, 0 \, E_{31}^T \, E_{12}^T \, 0 \, E_{32}^T \cdots E_{1l}^T \, 0 \, E_{3l}^T \cdots E_{1r}^T \, 0 \, E_{3r}^T]^T \in R^{n \times 3rn}$$

$$C_b \stackrel{\Delta}{=} [0 \, E_{21}^T 0 \, 0 \, E_{22}^T 0 \cdots 0 \, E_{2i}^T 0 \cdots 0 \, E_{2r}^T \, 0\,] \in R^{n \times 3rn}$$

$$\Delta(t) \overset{\Delta}{=} \begin{pmatrix} \Delta_{11} & 0 & 0 & & & & & & \\ 0 & \Delta_{21} & 0 & & 0_{3\times3} & & & 0_{3\times3} & \\ 0 & 0 & \Delta_{31} & & & & & & \\ & & & \Delta_{12} & 0 & 0 & & & \\ & 0_{3\times3} & & 0 & \Delta_{22} & 0 & & \vdots & \\ & & & 0 & 0 & \Delta_{32} & & & \\ & & & & & & \Delta_{1r} & 0 & 0 \\ & 0_{3\times3} & & & \cdots & & 0 & \Delta_{2r} & 0 \\ & & & & & & 0 & 0 & \Delta_{3r} \end{pmatrix} \in R^{3rn\times3rn}$$

The equation (5) is represented by the following generalized T-S cloud model plant, where $h_l(Z(t))$ is conveniently shown as $h_l$.

$$\begin{cases} \dot{x} = \sum_{l=1}^{r} h_l \cdot (A_l x + B_l u + D_{al}\omega) \\ z = C_a x + C_b u \\ y = \sum_{l=1}^{r} h_l \cdot (C_l x + D_{bl}\omega) \end{cases} \quad (9)$$

where $\omega = \Delta(t)\cdot z, \quad \Delta^T(t)\Delta(t) < I$.

With the equation (6), or dynamic compensator, the above equation is equal to the following one.

$$\begin{cases} \dot{x} = \sum_{l=1}^{r} h_l \cdot (A_l x + B_l \sum_{j=1}^{r} h_j \hat{C}_j x_c + D_{al}\omega) \\ \dot{x}_c = \sum_{l=1}^{r} h_l \cdot (\hat{A}_l x_c + \hat{B}_l \sum_{j=1}^{r} h_j (C_j x + D_{bj}\omega)) \end{cases} \quad (10)$$

From the equation (10), $\begin{pmatrix} \dot{x} \\ \dot{x}_c \end{pmatrix} = \sum_{l=1}^{r}\sum_{j=1}^{r} h_l h_j \left[ \begin{pmatrix} A_l & B_l\hat{C}_j \\ \hat{B}_l C_j & \hat{A}_l \end{pmatrix} \begin{pmatrix} x \\ x_c \end{pmatrix} + \begin{pmatrix} D_{al} \\ \hat{B}_l D_{bj} \end{pmatrix} \begin{pmatrix} \omega \\ \omega \end{pmatrix} \right]$ (11)

Introducing $X \overset{\Delta}{=} \begin{pmatrix} x \\ x_c \end{pmatrix}$, $A_{lj} \overset{\Delta}{=} \begin{pmatrix} A_l & B_l\hat{C}_j \\ \hat{B}_l C_j & \hat{A}_l \end{pmatrix}$, $B_{lj} \overset{\Delta}{=} \begin{pmatrix} D_{ai} \\ \hat{B}_l D_{bj} \end{pmatrix}$, $C_{lj} \overset{\Delta}{=} \begin{pmatrix} C_a & C_b\hat{C}_j \end{pmatrix}$, the state equation of close loop system with T-S cloud model is represented by the following.

$$\begin{cases} \dot{X} = \sum_{l=1}^{r}\sum_{j=1}^{r} h_l h_j \cdot (A_{lj} X + B_{lj}\omega) \\ z = \sum_{l=1}^{r}\sum_{j=1}^{r} h_i h_j \cdot C_{ij} X \end{cases} \quad (12)$$

**[Additional theorem]**

$\forall\, i$, $k_i > 0$, $\sum_{i=1}^{s} k_i = 1$, and $M_i$ is an optional equation $\rightarrow$

$$\sum_{i=1}^{s} k_i \cdot M_i \cdot M_i^T \geq \sum_{i=1}^{s}\sum_{j=1}^{s} k_i k_j \cdot M_i \cdot M_j^T \quad (13)$$

(the course of proof is not mentioned.)

**Theorem**

In order to the close loop system (12) is second order stable, it is necessary and sufficient that a positive definite matrix $P$ positive definite and satisfied by the following condition for $\forall\, i, j$ exists.

$$A_{ij}^T P + PA_{ij} + PB_{ij} B_{ij}^T P + C_{ij}^T C_{ij} < 0, \ \left(H_i \cap H_j \neq \phi\right) \qquad (14)$$

**(Proof)**

$$A = \sum_{i=1}^{r}\sum_{j=1}^{r} h_i h_j \cdot A_{ij}\ , \quad B = \sum_{i=1}^{r}\sum_{j=1}^{r} h_i h_j \cdot B_{ij}\ , \quad C = \sum_{i=1}^{r}\sum_{j=1}^{r} h_i h_j \cdot B_{ij}$$

According to the above equation, the equation (12) is equal to the following one.

$$\begin{cases} \dot{X} = AX + B\omega \\ z = CX \end{cases}$$

In order that this system is stable, it is necessary and sufficient that a positive definite matrix $P > 0$ given by $A^T P + PA + PBB^T P + C^T C < 0$ (15) exists from the theorem of bounded real value.

Then, the next results follow from that insert $A, B, C$ to the left side of the equation (15).

$$\begin{aligned}
&\left(\sum_{i=1}^{r}\sum_{j=1}^{r} h_i h_j \cdot A_{ij}^T\right)P + P\left(\sum_{i=1}^{r}\sum_{j=1}^{r} h_i h_j \cdot A_{ij}\right) + P\left(\sum_{i=1}^{r}\sum_{j=1}^{r} h_i h_j \cdot B_{ij}\right)\cdot \\
&\left(\sum_{i=1}^{r}\sum_{j=1}^{r} h_i h_j \cdot B_{ij}^T\right)P + \sum_{i=1}^{r}\sum_{j=1}^{r} h_i h_j \cdot C_{ij}^T C_{ij} \leq \left(\sum_{i=1}^{r}\sum_{j=1}^{r} h_i h_j \cdot A_{ij}^T\right)P + \\
&P\left(\sum_{i=1}^{r}\sum_{j=1}^{r} h_i h_j \cdot A_{ij}\right) + P\left(\sum_{i=1}^{r}\sum_{j=1}^{r} h_i h_j \cdot B_{ij} B_{ij}^T\right)P + \sum_{i=1}^{r}\sum_{j=1}^{r} h_i h_j \cdot C_{ij}^T C_{ij} \\
&= \sum_{i=1}^{r}\sum_{j=1}^{r} h_i h_j \cdot (A_{ij}^T P + PA_{ij} + PB_{ij} B_{ij}^T P + C_{ij}^T C_{ij})
\end{aligned}$$

In the derivate course of the above equation, consider the equation $\sum_{i=1}^{r} h_i = 1, h_i, h_j > 0$ to use the additional theorem.

For $\forall i, j$, when a positive definite matrix $P_j > 0$ given by the equation

$A_{ij}^T P + PA_{ij} + PB_{ij}B_{ij}^T P + C_{ij}^T C_{ij} < 0$, $(H_i \cap H_j \neq \phi)$, the close loop system follows from the equation (15) to be stable. **(The end of proof)**

It follows from the theorem that generalized plant of the $j^{th}$ rule should be stabilized by the compensator of all control rules that satisfy the equation $(H_i \cap H_j \neq \phi)$.

Therefore, the compensator of $j^{th}$ control rule should be designed that stabilizes generalized plant for all rules given by the equation $(H_i \cap H_j \neq \phi)$.

By using this result, the design approach of compensator of $j^{th}$ control rule follows from the additional theorem like that.

① For $\forall i$, determine the matrix $P_j > 0$ given by the equation

$$A_i P_j + P_j A_i^T + P_j C_{ai}^T C_{ai} P_j + D_{ai} D_{ai}^T - BB^T < 0, \quad (H_i \cap H_j \neq \phi) \qquad (16).$$

② For $\forall i$, determine the matrix $Q_j > 0$ given by the equation $N_{ij} = Q_j A_i + A^T Q_{j_i} + Q_j D_{ai} D_{ai}^T Q_j + C_{ai}^T C_{ai} - C^T C < 0, \quad (H_i \cap H_j \neq \phi)$

(17).

③ Check that satisfies the equation $\begin{pmatrix} P_j & I \\ I & Q_j \end{pmatrix} > 0$.

This is equal to calculation of symmetrical matrices $P_j > 0,\ Q_j > 0$ that satisfy the following LMI.

$$\text{For } \forall i, \begin{pmatrix} A_i P_j + P_j A_i^T + D_{ai} D_{ai}^T - BB^T & P_j C_{ai}^T \\ C_{ai} P_j & -I \end{pmatrix} < 0$$

$$\begin{pmatrix} Q_j A_i + A^T Q_{j_i} + C_{ai}^T C_{ai} - C^T C & Q_j D_{ai} \\ D_{ai}^T Q_j & -I \end{pmatrix} < 0$$

$$\begin{pmatrix} P_j & I \\ I & Q_j \end{pmatrix} > 0$$

Then, parameter matrix of $j^{th}$ compensator is obtained by the following equation.

$$\hat{A}_j = A_j + B_j C_{Cj} - B_{Cj} C_j + Q_j^{-1} C_{aj}^T C_{aj} - Q_j^{-1} N_{jj} (I - P_j Q_j)^{-1}$$
$$B_{Cj} = Q_j^{-1} C_j^T,$$
$$C_{Cj} = B_j^T Q_j (I - P_j Q_j)^{-1}$$

3. The off-line optimization of the design parameter of the triangle cloud control system by the parallel variable scaling rate chaos optimization approach

First, Using the Hybrid Chaos Optimal Method with the parallel variable chaos optimal method and the conjugation gradient method, the system configuration figure for optimizing the design parameters of the triangle cloud controller is shown fig.1.

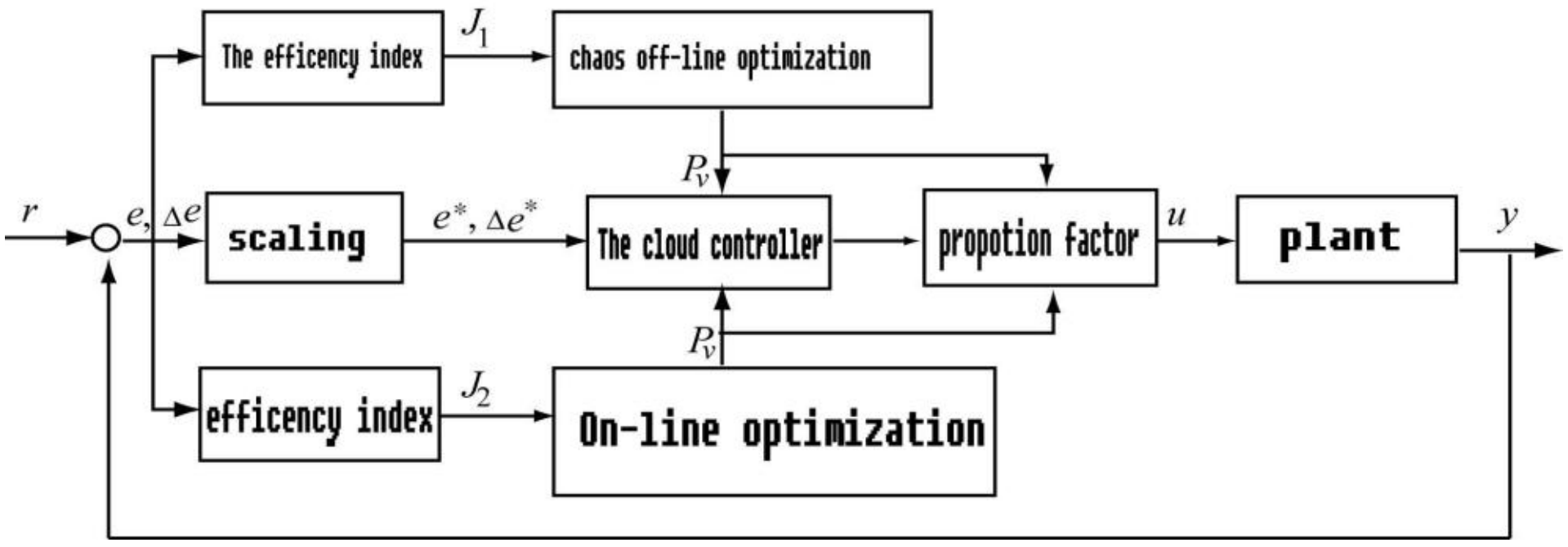

Fig.1 System configuration figure for design optimization of the cloud controller

As you're seen on figure, the triangle cloudy controller is a 2-input-1-output system construction and the front membership cloudy function is the triangle member cloudy function, the back member function is single point member cloud function.

Creating the expected value, variance, super variance, division number of the front member triangle cloudy function, and the expected value, division number and cloudy rule is decided on off-line that the efficiency index J1 gets the minimum with the parallel variable reducing rate chaos optimal method.

And then we decide the parameters of the cloudy controller on on-line that the efficiency index J2 gets minimum with the conjugation gradient method.

Using parallel variable reducing rate chaos optimization method, the efficiency index J1 is expressed like this.

$$J_1 = \min[\sum_{k=1}^{T} k^2 |e(k)| \Delta t]$$

This optimizes the control parameters with parallel variable scaling rate chaos optimal approach

$$J_1 = \min[\sum_{k=1}^{T} k^2 |e(k)| \Delta t] \quad (15)$$

This expression means that the error area gets minimal all over the control interval.

In the expression, $e(k) = r(k) - y(k)$ means error, $\Delta t$ means sampling interval, $r(k)$ means the quantity of the target institution, $y(k)$ is the output of the plant.

And then let express the parameters; $Ex_{x_1}, Ex_{x_2}$, - the expected value, $En_{x_1}, En_{x_2}$ -the variance, $He_{x_1}, He_{x_2}$ -the super variance of the front member cloud function and $Ex_u$ -the expected value of the after member cloud function, the cloud division number of the first input variable of the cloud controller, the cloud division number of the second input variable , the division number of the after member cloud function, creating the cloud rules and the proportion coefficients as $P_v, v = 1,2,\cdots$ that is $P_1 = Ex_{x_1}, P_2 = Ex_{x_2}\ P_3 = En_{x_1}, P_4 = En_{x_2},\ P_5 = He_{x_1}, P_6 = He_{x_2},\ P_7 = Ex_u, P_8 = m_1, P_9 = m_2,$ $P_{10} = o, P_{11} = RL,\ P_{12} = K_u$

Here $P_1, P_{2,}\ P_{3,}\ P_4, P_5, P_6, P_7$ are the parameters of the front and the after member cloud function so in the case of the division number is $P_{5,}\ P_6, P_7$ , they become as

$P_1^{\ i}, P_3^{\ i},\ P_5^{\ i}, P_2^{\ j}, P_4^{\ j}, P_6^{\ j},\ P_7^{\ h},\ i = m_1,\ j = m_2,\ h = o$ and

$P_8 = RL = [RL^1, \cdots, RL^\kappa, \cdots],\ \kappa = m_1 \times m_2$.

And then $P_1, P_2, P_7$ have the value of from-1 to 1, $P_3, P_4, P_5, P_6$ have the value of from 0 to 1, $P_8, P_9, P_{10}$ have the value of 1 to 20 , $P_{11}$ has the value of 1 to o which is the division number of the after member cloud function, $P_{12}$ has the value of $P_u = \max(|u_{\min}|, |u_{\max}|)$ .

Now let's study about Logistic mapping- $\alpha_{s+1} = 4\alpha_s(1-\alpha_s),\ \ s = 1,2,\cdots,N,\ \alpha_0 \in (0,1)$

to use the parallel variable scaling chaos optimal approachss

In this expression N is the degree of the chaos optimization.

The chaos variable $\alpha_S$ takes the value of from 0 to 1, so we must modify $P_v$ as the following expression to determine the parameters of the cloud controller.

$$P_1, P_2, P_7 = -2\alpha + 1$$
$$P_3, P_4, P_5, P_6 = \alpha$$
$$P_8 = P_9 = P_{10} = round(20 \times \alpha)$$
$$P_{11} = round(o \times \alpha)$$
$$P_{12} = round(P_u \times \alpha) \qquad (3)$$

The optimal process for which determine the parameters of the triangle cloud controller is following.

Step1. In the cloud controller, the number of parameter which will be determined is following.

$$\gamma = m_1 \times 3 + m_2 \times 3 + o + 4 + m_1 \times m_2$$

Take the different forty-one initial random values in the interval between 0 and 1.

Step2. The initial values that are gotten are substituted to expression (2) and then we can get the $\gamma$ chaos trajectory variables.

Step3. Using expression (3), calculate the control force by substituting modified $\gamma$ variable to the output cloud controller and then calculate the efficient index J1 based on the expression (1) by substituting to the system model.

Step4. If the termination condition is satisfied, you should complete searching and go to step5, but if it isn't, you should return to setp2.

Step5. Look for the minimal value of the system efficient index and

Put out the optimal solution at that time.

## 4. On-line optimization of the design parameter of the Triangle Cloud Controller with the Conjugation Gradient Approach

If you optimize the parameters of the controller on off-line with the previous chaos search approach, you couldn't react actually about the random parameter variance.

So we can determine the parameters approximately with the chaos search approach and should optimize the parameters of the controller on on-line with the gradient approach by using them as initial value.

For this, we should decide the efficient index as the following expression.

$$J_2 = \frac{1}{2} e(t)^2, \quad e(t) = r(t) - y(t) \qquad (4)$$

The negative gradient direction of the parameter vector about efficient index J2

$\mathrm{P} = [\mathrm{P}_1, \mathrm{P}_2, \mathrm{P}_3, \mathrm{P}_4, \mathrm{P}_5, \mathrm{P}_6, \mathrm{P}_7, P_8, P_9, P_{10}, \mathrm{P}_{11}, P_{12}]^{\mathrm{T}}$ is

$$W' = -\nabla J_2(P) = \sum_{k=1} e_k \frac{\partial y_k}{\partial P_k} = \sum_{k=1} (r_k - y_k) \frac{\partial y_k}{\partial P_k} \qquad (5)$$

And the step width $\eta$ of the optimal search direction is determined as the following expression.

$$\frac{dJ_2(x_{i,k} + \eta W'_k)}{d\eta} = 0 \qquad (6)$$

The correction expression of the parameter vector which we are going to obtain is

$P_{\vartheta,k+1} = P_{\vartheta,k} + \eta_k \cdot W_k'$.

Here, $P_{i,k}$ is the ith parameter vector in the kth iteration, $W_k'$ is the search direction vector in the kth search.

The algorithm of the conjugation gradient is

$$W_k' = -\nabla J_2(P_{i,k}) + \delta_{k-1} \cdot W_{k-1}' \qquad (7)$$

And the conjugation coefficient of the (k-1)th step is

$$\delta_{k-1} = \frac{[\nabla J_2(P_{\vartheta,k})]^T \cdot [\nabla J_2(P_{\vartheta,k})] - [\nabla J_2(P_{\vartheta,k})]^T \cdot [\nabla J_2(P_{\vartheta,k-1})]}{[\nabla J_2(P_{\vartheta,k-1})]^T \cdot [\nabla J_2(P_{\vartheta,k-1})]}.$$

The process for which determines the parameters of the cloud controller with the conjugation gradient is following.

Step1. The $\gamma$ optimal solutions which is gotten by searching with the parallel variable scaling chaos optimal approach is substituted to the expression (5) as the initial value.

Step2. Calculate the step width $\eta$ of the optimal search direction by using expression (6).

Step3. Calculate the efficient index J2 by substituting the new $\gamma$ variables as the parameters of the cloud controller, which are calculate by being based on expression (7).

Step4. Substitute $P_{\vartheta,k+1}$ as $P_{\vartheta,k}$ and return to step1.

## 5. Experiment

Now, let’s see the plant is express as the following discrete system equation.

$$y(k) = 3.737y(k-1) - 4.212y(k-2) + 1.492y(k-3) + 0.17u(k-1) -$$
$$-0.238u(k-2) + 2.94u(k-3) + \delta(k)$$

Here $\delta(k)$ is the white noise whose mean value is zero and variance is 1.

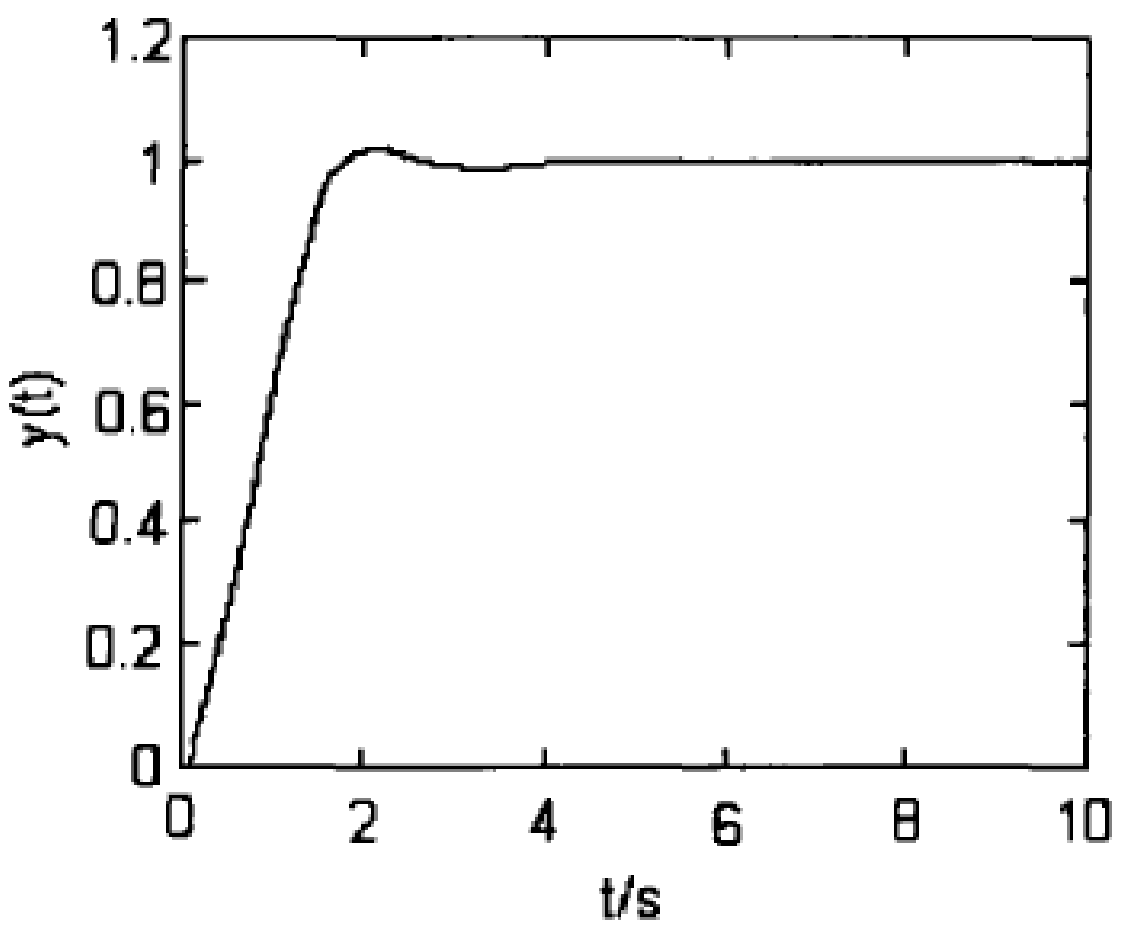

Fig.2 The response curve line of the system

First, Fig2 shows the result that are obtained by searching the global optimal solution accorded with the efficient index J1 on 837 steps with the chaos optimal approach and then obtaining the solution accorded with the efficient index J2 on 261 steps with the gradient approach and finally doing cloud control.

When fix the search completion as $J = 10^{-3}$ and optimize the cloud controller with the single chaos optimal method, the conjugation gradient descent method, genetic algorithm and the hybrid chaos optimal method, the frequency of obtaining the optimal solution is following.

Table. The degree of obtaining the optimal solution with the different optimal methods

| No. | The optimization approach | The degree of finding the optimal solution |
|---|---|---|
| 1 | The single chaos optimization method | 1539 |
| 2 | The conjugation gradient descend method | $\infty$ |
| 3 | Genetic algorithm | 1651 |
| 4 | The hybrid chaos optimization method | 1098 |

From the table, the times on which obtain the optimal solution with the conjugation gradient is $\infty$ so this method is easy to fall into the limit minimum value, you could notice that the global optimal solution can’t be obtained.

And you can notice that the single chaos optimal method and the genetic algorithm are the global chance search method so we can find the global solutions with them and although the single chaos optimal method is more predominate than the genetic algorithm, they are not so good than the proposed hybrid chaos optimal method.

You can also notice that this regulator has a disturbance restraint character from the system response trajectory.

**Conclusion**

First, the approach which optimizes the design parameters of the triangle cloud control system on off-line with the parallel variable scaling chaos optimization method had been proposed.

Next, the approach which optimizes the design parameters of the triangle cloud control system with the gradient method had been proposed. Next, the efficiency of the proposed approaches had been verified through the computer simulation experiment.